# Modulating nano-inhomogeneity at electrode-solid electrolyte interfaces for dendrite-proof solid-state batteries and long-life memristors

Ziheng Lu[a,1,*], Ziwei Yang[a,1], Cheng Li[a], Kai Wang[a], Jinlong Han[a], Peifei Tong[a], Guoxiao Li[a], Bairav Sabarish Vishnugopi[b], Partha P. Mukherjee[b], Chunlei Yang[a,*], Wenjie Li[a,*]

[a] *Shenzhen Institutes of Advanced Technology, Chinese Academy of Sciences, Shenzhen 518055, Guangdong, China*
[b] *School of Mechanical Engineering, Purdue University, West Lafayette, Indiana 47907, United States*

[*]Corresponding authors. E-mail addresses: zh.lu1@siat.ac.cn (Z. Lu); cl.yang@siat.ac.cn (C. Yang); wj.li@siat.ac.cn (W. Li);

**Abstract**

The penetration of dendrites in ceramic lithium conductors severely constrains the development of solid-state batteries (SSBs) while its nanoscopic origin remain unelucidated. We develop an *in-situ* nanoscale electrochemical characterization technique to reveal the nanoscopic lithium dendrite growth kinetics and use it as a guiding tool to unlock the design of interfaces for dendrite-proof SSBs. Using $Li_7La_3Zr_2O_{12}$ (LLZO) as a model system, *in-situ* nanoscopic dendrite triggering measurements, *ex-situ* electro-mechanical characterizations, and finite element simulations are carried out which reveal the dominating role of $Li^+$ flux detouring and nano-mechanical inhomogeneity on dendrite penetration. To mitigate such nano-inhomogeneity, an ionic-conductive homogenizing layer based on poly(propylene carbonate) is designed which *in-situ* reacts with lithium to form a highly conformal interphase at mild conditions. A high critical current density of 1.8mA cm$^{-2}$ and a low interfacial resistance of 14Ω cm$^2$ is achieved. Practical SSBs based on LiFePO$_4$ cathodes show great cyclic stability without capacity decay over 300 cycles. Beyond this, highly reversible electrochemical dendrite healing behavior in LLZO is discovered using the nano-electrode, based on which a model memristor with a high on/off ratio of ~10$^5$ is demonstrated for >200 cycles. This work not only provides a novel tool to investigate and design interfaces in SSBs but offers also new opportunities for solid electrolytes beyond energy applications.

**Introduction**

Lithium-ion batteries (LIBs) since their commercialization have revolutionized the energy storage sector and are presently ubiquitous across portable electronics. However, recent advancement of the electric vehicle industry and grid-scale storage necessitate energy storage solutions beyond current LIB technology with aggressive demands on safety, price, and energy density. [1] Especially, in order to reach the target of 500 Wh kg$^{-1}$, replacing the current graphite anode with lithium metal has been considered critical. [2-5] Unfortunately, lithium (Li) metal is highly reactive and tends to electrodeposit into a dendritic morphology during cycling, which inevitably paves the way for cell failure. Extensive research efforts have been made in this regard and various strategies such as tuning the electrolyte composition, [6, 7] coating the Li metal with artificial layers, [8, 9] designing porous current collectors, [10] and utilizing solid electrolytes (SEs) [11, 12] have been proposed. Amongst these, replacing the conventional liquid electrolyte with a mechanically stiff and Li$^+$-conductive SE has been regarded as one of the most promising solutions to address both the safety concerns and energy density limitations of LIBs. On one hand, SEs are less flammable when compared to the organic liquid electrolyte. On the other hand, due to their mechanical rigidity, SEs are also expected to suppress the growth of uneven Li deposits.. [13]

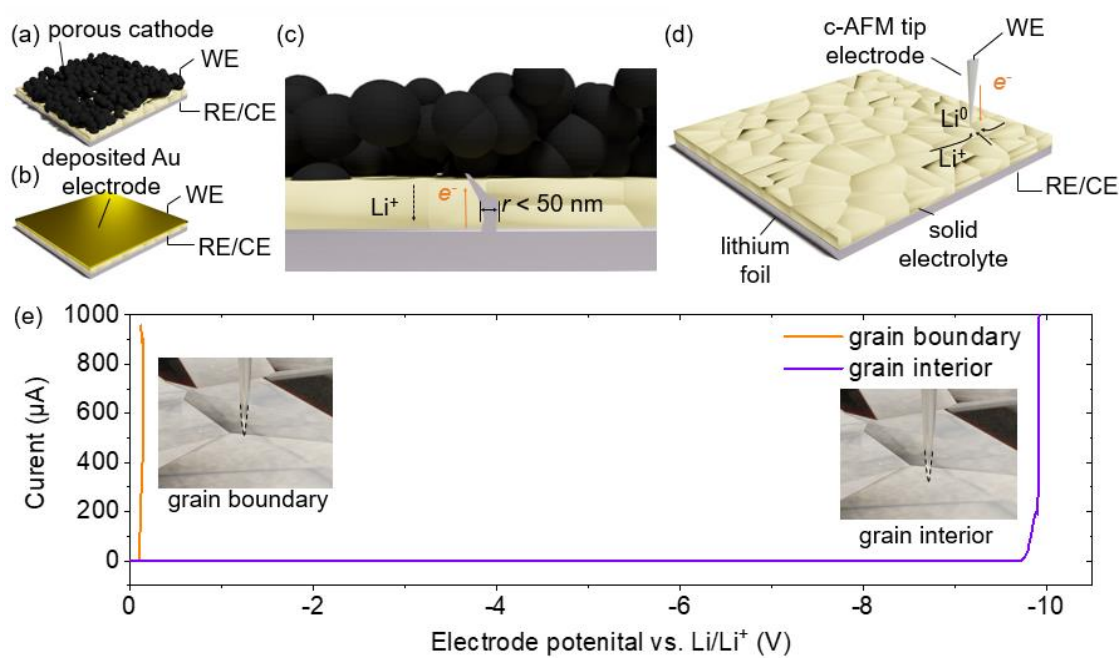

**Figure 1** Illustration of the macroscopic electrochemical measurement setups using (a) porous cathodes and (b) deposited Au electrodes. Illustration of (c) The penetration of lithium filaments in a conventional macroscopic electrochemical measurement setup. (d) Illustration of the nanoscale electrochemical measurement setup used in the current study where the c-AFM tip is used as the WE. (e) Typical current response of applying a varying bias at grain boundaries and grain interiors using the c-AFM nano-electrode.

Among the SEs that have been discovered, ceramic oxides are one of the most promising families due to their high ionic conductivity, excellent fire redundancy, good mechanical strength, and the ease of fabrication in air. In particular, the lithium-stuffed garnet $Li_7La_3Zr_2O_{12}$ (LLZO) has garnered tremendous research attention since its discovery in 2007. [14] When properly doped, it displays a high ionic conductivity of over $10^3$ S cm$^{-1}$ and a Young's modulus of ~150 GPa. [15, 16] In addition, LLZO is one of the very few SEs that are chemically stable toward Li metal. [15, 16] Owing to these attributes, LLZO has also been regarded as a promising candidate for employment in solid-state lithium-metal batteries (SSLMBs). In fact, one of the rationales behind the early investigation of LLZO, inferred from the linear elasticity model by Monroe and Newman [17] was that its high shear modulus could suppress the growth of Li dendrites. However, this hypothesis has recently been challenged by several experimental findings that reveal the penetration of Li metal in LLZO and observe internal short circuit at high current densities. [18] Since then, significant research efforts have been placed in understanding the mechanisms underlying the penetration behavior of Li metal. Hu *et al.* found that LLZO is poorly wetted by Li, even in its molten state. The poor interfacial contact leads to high local current densities and triggers local "hot spots" for Li to penetrate. [19, 20] Sharafi and co-workers confirmed that the poor wettability is due to the surface contaminants such as LiOH and $Li_2CO_3$. [21] To quantify the capability of LLZO in blocking the dendrites, they further proposed the concept of critical current density (CCD) which is defined as the highest current density before a Li|LLZO|Li cell shorts. [22] Usually, for an LLZO pellet with untreated surface, the CCD does not exceed 0.2 mA cm$^{-2}$. This value is significantly lower than the 2 mA cm$^{-2}$ threshold needed for practical applications. Apart from the contact issues, local inhomogeneity has also been identified as a source of instability for Li deposition. Importantly, the grain boundary is deemed to be a preferential nucleation site due to its high electron conductivity, [23] low elastic modulus, [24] and the low fracture toughness. [25-27] Recently, Porz *et al.* proposed a Griffith-like crack extension model where lithium metal infiltrates into the electrolytes through surface defects. [27] Considering the fundamental difference between the Li penetration in ceramic SEs and its dendritic-like growth in liquid electrolytes, we will use the term "filament" instead of "dendrite" throughout the article. In literature, various strategies have been proposed to enhance the stability of Li|LLZO interfaces and improve the CCD values, including physical/chemical treatment of the interfaces to enhance the interface contact, [19, 20] designing composite anodes to modify its interactions with LLZO, [28, 29] and modifying grain boundary properties using sintering aids. [30] Aided by these recent advancements, the CCD has been reported to reach ~1 mA cm$^{-2}$. [31-34] Despite the significant efforts laid towards understanding and mitigating the growth of filaments in LLZO, to date, its electro-chemo-mechanical origin is yet to be revealed in detail. Especially, almost all previous studies have been based on macroscopic electrochemical measurements as shown in **Figure 1(a)** to **(b)**, which lack the spatial resolution to directly measure the triggering conditions and growth kinetics of an individual filament due to its small size as illustrated in **Figure 1(c)**. Moreover, it has been argued that Li filament prefers to penetrate the grain boundaries instead of grain interiors.[18] However, at what conditions such preferential nucleation happens and how it can be quantified still remains to be answered. In addition, the nanoscopic driving force of such inhomogeneity is still unresolved. Understanding such nanoscale interactions is also critical for the design of interface layers and prevention of filament growth.

In this work, we develop a high-resolution *in-situ* characterization technique to probe the local dynamics and the electro-mechanical origin of lithium filament penetration, and use it as a guiding tool to design a highly efficient interphase to prevent short circuit and to achieve stable deposition. We exploit the extreme spatial resolution of conductive-atomic force (c-AFM) and utilize the AFM tip as the working electrode to selectively trigger dendrites in LLZO as illustrated in **Figure 1(d)**. By applying electric biases on the tip of the c-AFM, we quantitatively measure the electrochemical responses of the lithium plating processes and the filament growth kinetics with the resolution down to nanometers. In particular, the intrinsically different responses of grain interior and grain boundaries to electrochemical lithium deposition are revealed for the first time. We find that the critical electrical bias to induce lithium filament growth at the grain boundary is ~1/100 of that in the grain interior, as shown in **Figure 1(e)**. Such a striking difference points to the fact that the nanoscale inhomogeneity of the LLZO surface results in weak spots which enables a preferential penetration pathway for lithium metal. Further *ex-situ* nano-electro-mechanical AFM characterizations and finite element simulations suggest that the detouring and concentration of $Li^+$ flux at the interface between Li and LLZO grain boundaries is the major contributor that triggers penetration of the Li filament. Built on this understanding, we develop a highly efficient filament-proof interphase based on

poly(propylene carbonate) (PPC) that is able to homogenize the local Li$^+$ flux and increase the CCD. Such an interphase *in-situ* reacts with lithium metal at mild conditions and forms a highly conformal interface. The interfacial resistance drops from ~1000 Ω cm$^2$ to an exceedingly low 14Ω cm$^2$. The CCD value also increases from ~0.2 mA cm$^{-2}$ to 1.8 mA cm$^{-2}$ which is very close to the requirement of SSBs under practical conditions. Full SSLMBs are demonstrated using LiFePO$_4$ (LFP) and LiNi$_{0.5}$Co$_{0.2}$Mn$_{0.3}$O$_2$(NCM523) as cathodes. Both show excellent stability up to 300 cycles. Apart from the application in SSBs, we also discover signature memristive switching characteristics of filaments in LLZO under cyclic conditions, which is essential in neuromorphic computing and non-volatile memory devices. [35, 36] A model memristor is designed and demonstrated based on the nano-electrode with unprecedented stability of over 200 cycles and an on/off ratio up to 10$^5$. The novel characterization technique developed in this work not only facilitates the understanding and the design of highly efficient interfaces that can potentially unlock the capabilities of SSLMBs, but also opens up new opportunities for solid electrolytes beyond energy applications. [35, 36]

**Results and Discussions**

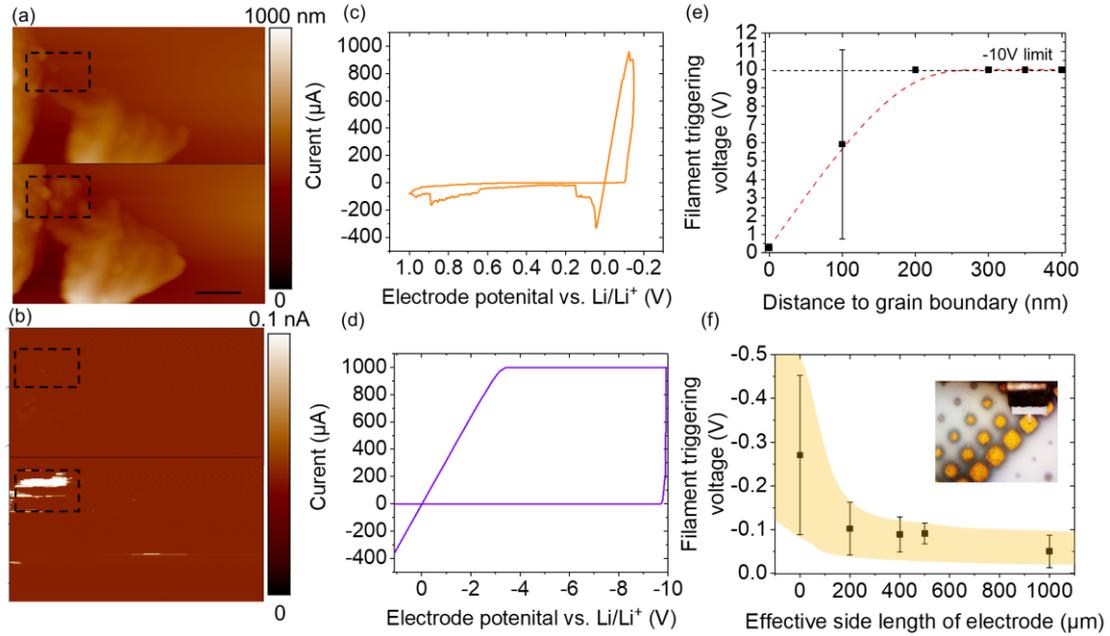

**Figure 2** (a) The morphology change of the LLZO surface before (upper panel) and after (lower panel) the penetration of lithium. (b) The current map of the LLZO surface using c-AFM by applying a +0.1V bias before (upper panel) and after (lower panel) the penetration of lithium. The scale bars in (a) and (b) are 200nm. Typical current response of applying a cyclic bias at (c) grain boundaries and (d) grain interiors using the c-AFM nano-electrode. The filament triggering bias measured (e) with c-AFM tips at different distances to the grain boundaries and (f) with deposited Au electrodes with different sizes. The inset in (f) shows the digital image of the deposited Au electrodes with different sizes.

The experimental setup for the nano-electrochemical characterizations is shown in **Figure 1(d)**. The c-AFM tip is used as the working electrode (WE) whereas Li metal functions as both the counter (CE) and the reference electrode (RE). The working principle of the system is illustrated in **Figure S1** and explained in detail in the supplementary information (SI). By applying a reductive bias on the WE (*i.e.*, negative with respect to Li/Li$^+$), the Li$^+$ ions are drawn from the LLZO to the tip of the c-AFM and are reduced to form lithium metal (Li$^0$). In principle, when the bias is small, the current is solely contributed by the flow of Li$^+$ ions and can be used as proxy for the deposition rate of lithium metal. When the electric bias gets large enough, the lithium metal can penetrate through the LLZO pellet causing short circuits. Under such circumstances, the measured current goes through an abrupt change, indicating the transition from ionic to electronic conduction. In this study, we aim to monitor the electrochemical conditions (*i.e.*, the bias and the current density) at which this transition happens. We use this value, *i.e.*, the triggering bias, as the major proxy to quantify how easy it is for lithium filaments to penetrate through the LLZO as illustrated in **Figure 1(e)** and **Figure S1**. In particular, for polycrystalline SEs such as LLZO, we focus on understanding how the grain interior differs from the grain boundaries as the latter has been speculated to be a weak spot for the penetration of lithium filaments, while actual quantitative electrochemical measurements have not been reported. The typical electrochemical results of a grain boundary are shown in **Figure 1(e)**. The current does not show significant increase until the bias is larger than -0.12V *vs*. Li/Li$^+$ where a surge in the conductivity is detected. Such a surge agrees with the metallic filament penetration and is further supported by a back scan where an almost perfect linear relation between the current and the voltage is observed, indicating its pure ohmic nature, as shown in **Figure 2(c)**. [27, 37] Further mapping of the morphology change and the corresponding conductivity also confirms such a result. As shown in **Figure 2(a)** and **(b)**, highly conductive bumps emerge after the nano-electrochemical measurements. These bumps are essentially the penetrated lithium dendrites and the points of short circuits. In contrast, for grain interiors, despite the high number of trials, we were only able to induce the filament growth for very limited times and all of them are triggered by large biases. A typical case shown in **Figure 1(e)** and **Figure 2(d)** where penetration of the metallic filament happens at < -9 V *vs.* Li/Li$^+$. Such an astonishing 100-fold increase in the triggering bias for filament penetration unambiguously points to the fact that the grain boundary acts as a weak spot in the electrolyte and its implications should be mitigated as much as possible. In fact, we carried out multiple independent tests and gathered the statistics. As shown in **Figure 2(e)** and **Figure 2(f)**, as we varied the distance of the c-AFM electrode to the grain boundary and changed the area of the electrode, the filament triggering bias for grain interior is consistently larger than the grain boundary. Therefore, though LLZO can be intrinsically capable of blocking lithium filaments, its filament-proof capability can be unlocked only when such interfacial nano-inhomogeneities are eliminated.

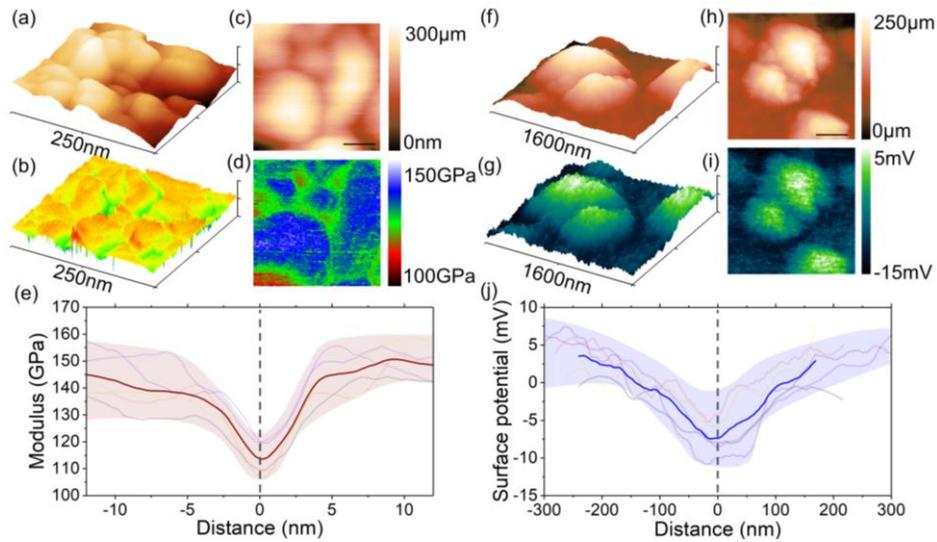

**Figure 3** (a) The morphology and (b) the modulus mapping of the LLZO surfaces. (c) and (d) are the 2D view of selected areas. The scale bars in (c) and (d) are 50 nm. (e) The local modulus as a function of the distance to the grain boundary. (f) The morphology and (g) the surface potential mapping of the LLZO surfaces. (c) and (d) are the corresponding 2D views. The scale bars in (c) and (d) are 500 nm. (e) The surface potential as a function of the distance to the grain boundary. For (e) and (j), the brown and blue bold line in the foreground is the average of the results from 5 separate tests at diffract locations as indicated by the lines in the background. The shaded areas are guide to the eye to indicate the error range.

Considering the hundred-fold weakening of the SE at the grain boundary against lithium penetration, we further explored its mechanical and the electrical origin and use it to guide the design of an efficient filament-proof interphase. **Figure 3(a)-(e)** shows the nano-mechanical measurement of a typical LLZO surface. The mechanical stiffness is clearly lower at the grain boundaries compared with the grain interiors. In order to avoid the influence of the abrupt changes of surface morphology on the accuracy of elastic property measurement, we specifically chose a relatively flat area and show the results in **Figure 3(d)** and **(e)**. The decrease at the grain boundaries can be as high as 30% compared with the interior. This is further supported by sampling of the moduli along the vertical lines to the grain boundaries, see **Figure 3(f)**. By averaging 5 individual measurements, the moduli of LLZO decrease from ~145GPa in grain interiors to ~120 GPa at grain boundaries. Such results are in good agreement with the simulations done by Yu *et al*. [24] It is worthwhile to note that, despite the significant drop in the elastic moduli, the value is still significantly higher compared to lithium metal. Therefore, the elastic softening should not be the sole factor that governs the filament growth. However, such mechanical inhomogeneity may serve as an initiator for the preferential lithium deposition and assists the Griffith-like crack extension mechanism proposed by Porz *et al*. [27] Beside the mechanical origin, the electronic aspects have also been speculated to affect the lithium deposition stability. [13][23, 38-40] Here, we noticed that due to the electron insulating nature of LLZO, local inhomogeneity of electrical potential may build up at the LLZO|electrode interfaces. Such variance of the electric local electric potential at the interface may give rise to nucleation preferences for lithium. In fact, at grain boundaries, it is known that certain charged depleted or enriched region exist (also known as the space charge layer/space charge region). [41, 42] We measure such local variance of electric potential by mapping the surface potential to the morphology using the Kelvin probe force microscopy. [43] The results are shown in **Figure 3(f)** to **(j)**. It is found that the surface potential of LLZO at grain boundaries tends to decrease by ~10 mV compared with the grain interiors, see

**Figure 3(j)**. Such a decrease corresponds to an e$^-$ accumulation or Li$^+$ depletion and is in agreement with a very recent MD simulation by Shiiba *et al*.[44] The decrease of Li$^+$ concentration may result in lowered conductivity and therefore a preference of lithium extrusion.[45],[46] To study how the variation of local Li$^+$ diffusion could result in lithium penetration, we simulated the Li$^+$ flux in a polycrystalline LLZO with grain sizes on the order of ~10μm, see **Figure S2-S3** and section 2 of SI for details. As shown in **Figure 4(a)** and **(e)**, due to the low ionic conductivity of the grain boundaries, significant detour of Li$^+$ flux is observed. To minimize the probability of travelling through grain boundaries, Li$^+$ prefers to transport through certain grains instead of the others. This leads to strong spatial variation of current densities in LLZO and such inhomogeneity extends to the Li|LLZO interface. In fact, at the junction between Li and LLZO, the maximum current density within certain grains is over 10 times larger than the average value, as shown in **Figure 4(f)**. It is worthwhile to note that such a mechanism of current concentration would occur even under an ideal LLZO-Li contact. It is fundamentally different from the previously studied scenario, which involves imperfect contact due to limited interfacial wetting.[21] Essentially, in polycrystalline LLZO, even if the contact between LLZO and Li is perfect, "hot-spots" with high current densities still exist. Therefore, to avoid such "hot-spots", an interlayer with homogeneous ionic conductivity is necessary, as illustrated in **Figure 4(c)** and **(d)**. The effect of inserting such a homogenizing layer (H.L.) is simulated in **Figure 4(b)** where a grain boundary-free solid electrolyte with an ionic conductivity of 10$^{-4}$ S cm$^{-1}$ is attached between Li and LLZO. With such a layer, the Li$^+$ flux is effectively smoothened and the current

density at the Li|H.L. interface shows almost no fluctuation as shown in **Figure 4(f)**. Interestingly, such a homogenizing effect is relatively strong and the current density spike quickly drops to the average value when the H.L. is ~3μm thick. This value provides us with a guiding principle for the design of the interlayer.

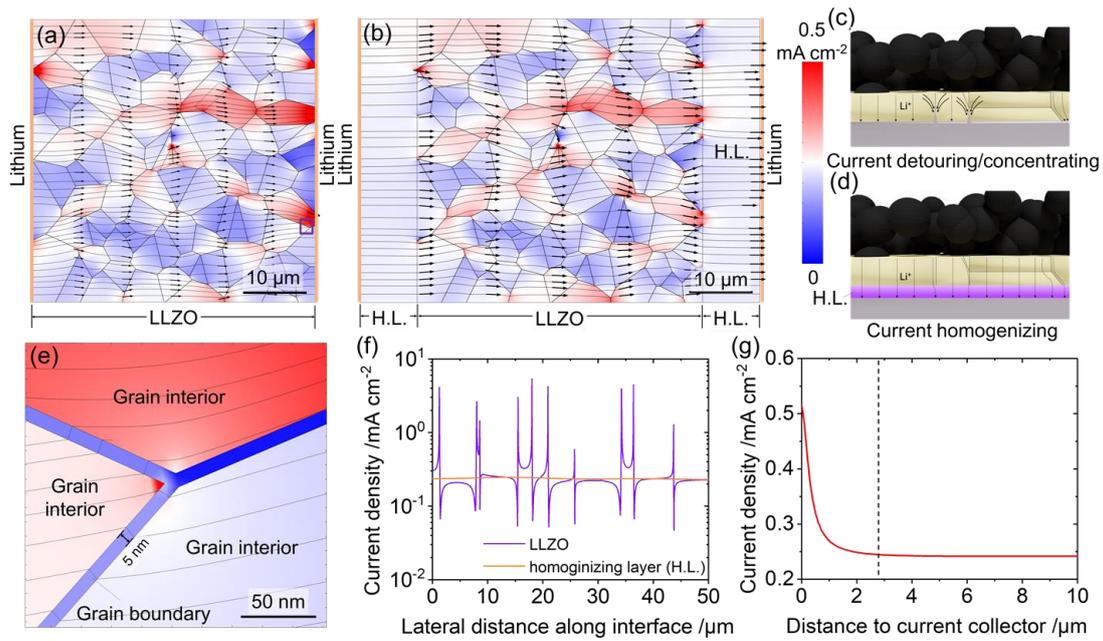

**Figure 4** Li$^+$ flux distribution in (a) polycrystalline LLZO and (b) H.L.|LLZO|H.L. Illustration of (c) the current detouring and concentrating at intrinsic Li|LLZO interfaces and (d) the homogenizing effect at the H.L. modified interfaces. (e) Detailed Li$^+$ current density distribution at grain boundaries. This is a zoomed image of the region highlighted with purple box in (a). (f) The current density at Li|LLZO interfaces and the corresponding value after inserting H.L. (g) The maximum current density in H.L. with respect to the distance to the LLZO|H.L. interface.

Based on previous simulation results, an ideal homogenizing layer is required to have the following characteristics: 1) It should display no spatial inhomogeneity in ionic conductivity. 2) The thickness should be on the order of μm and the ionic conductivity should be close to $10^{-4}$ S cm$^{-1}$. 3) It should be able to form good contact with both Li and LLZO and display small interfacial resistances. Considering these guidelines, we developed a novel polymeric interphase based on PPC. We chose this material because it has been proposed as a solid electrolyte when combined with lithium salt and shows descent ionic conductivity. [47-49] More importantly, PPC goes through a catalytic de-polymerization reaction when heating together with Li. [50-52] Such a de-polymerization reaction can facilitate the contact between PPC and Li. Therefore, the handling of highly hazardous molten lithium is no longer necessary. **Figure 5(a)** shows the surface morphology of LLZO surface covered by the PPC homogenizing layer. Compared with the untreated LLZO surface which is shown in **Figure 5(b)**, it is significantly smoother which indicates the spatial homogeneity. **Figure 5(c)** and **(d)** show the cross-section of the Li|H.L.|LLZO interface. The contacts between the homogenizing layer and the two ends are good. To verify the capability of achieving the homogenizing effect after inserting such a layer, we conducted nano-electrochemical tests using c-AFM. The surface morphology of LLZO and H.L is shown in **Figure 5(e)** and **(f)** and the corresponding roughness is calculated in **Figure 5(g)**. Morphologically, the H.L. layer covers the grain boundaries and other surface defects on LLZO, smoothing out the local variance. The electrochemical response of the two surfaces is shown in **Figure 5(h)** with the typical I-V curves showcased in **Figure 5(i)**. The filament triggering bias on the H.L. surface is not only significantly larger than that of the LLZO surface (grain boundary regions), but also very consistent throughout the entire region. Such a result indicates the PPC-based H.L. should be effective in enhancing the electrodeposition stability at the interface and preventing the growth of lithium filaments.

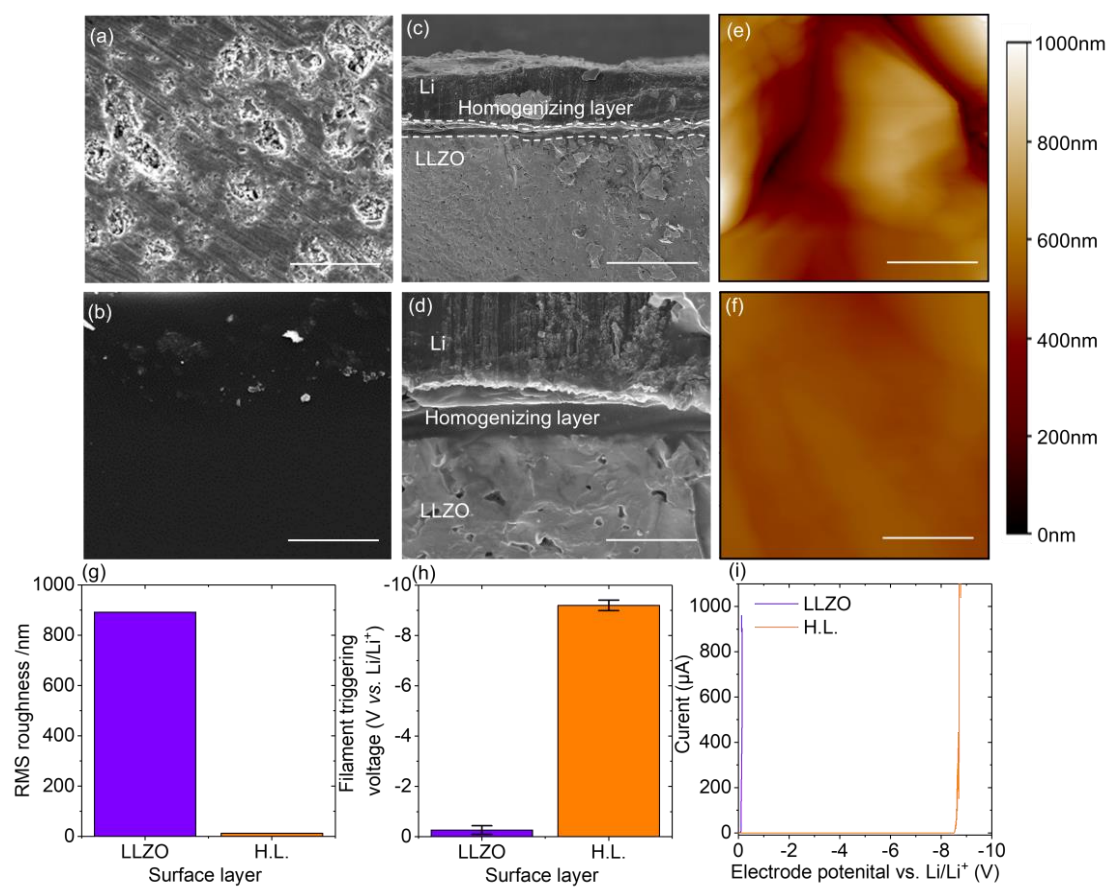

**Figure 5** The morphology of (a) polished LLZO surface and of (b) H.L. The scale bars in (a) and (b) are 50μm. (c)(d) The cross-section of Li|H.L.|LLZO interfaces. The scale bars in (c) and (d) are 100μm and 50μm, respectively. The morphology (e) of a fractured LLZO pellet showing its grain boundaries and (f) of a H.L characterized using AFM. The scale bars in (e) and (f) are (g) 2μm. The root mean square (RMS) roughness of the two surfaces. (h) The filament triggering bias on LLZO and H.L. and (i) the typical electrochemical responses.

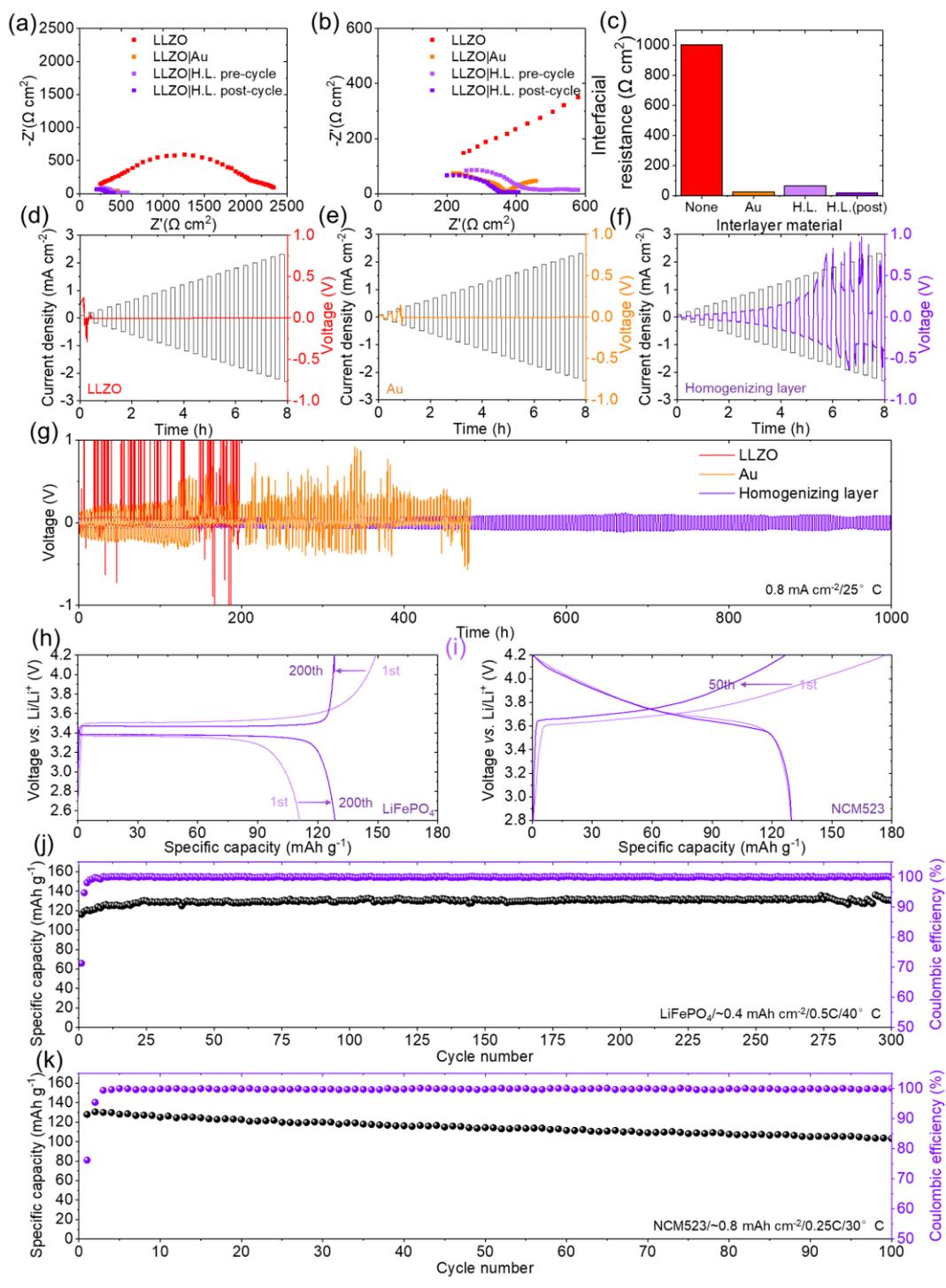

**Figure 6** (a) The EIS of Li|LLZO|Li cells with no interlayers, Au interlayers, and H.L. (b) The enlarged image of (a). (c) The interfacial resistance between Li and LLZO with different interlayers. The current density response of Li|LLZO|Li cells with (a) no interlayer, with (b) Au interlayer, and with (c) H.L. under increasing current densities.

(d) Voltage response of Li|LLZO|Li cells with different interlayers under cyclic galvanostatic currents. The typical charge-discharge curve of an SSB using (h) LFP and (j) NCM523 as cathode. The specific capacity of SSBs using (j) LFP and (k) NCM523 as cathode under cyclic conditions.

The practical macroscopic electrochemical performance of the H.L. interphase is further evaluated using Li|LLZO|Li cells. The electrochemical impedance spectroscopy (EIS) results are shown **Figure 6(a)-(b)**. Such an interphase is effective in reducing the interfacial resistance between Li and LLZO. The area resistance is ~50Ω cm$^2$ after inserting the PPC-based H.L. while the value for the intrinsic LLZO|Li interface without modification is ~1000Ω cm$^2$, as shown in **Figure 6(c)**. Such a significant drop in impedance comes from the highly conformal interphase formed during the *in-situ* de-polymerization process. Interestingly, we notice that after cycling the cell, the impedance decreases further to ~14Ω cm$^2$, such a value is even smaller than the pure inorganic Au|LLZO interface. This is probably because de-polymerization of PPC continued during cycling and further increased the conductivity of the interlayer. The H.L. not only significantly reduces the interfacial resistance, but also enlarges the CCD, which is one of the major bottlenecks for a practical SSB. As shown in **Figure 6(d)-(e)**, the CCD for the intrinsic LLZO|Li interface is <0.1 mA cm$^{-2}$ while for the Au-modified case is ~0.2 mA cm$^{-2}$. In comparison, in the H.L.-inserted case, the CCD reaches 1.8 mA cm$^{-2}$, almost 10 times higher than the Au-modified case. Such a drastic increase in the capability of blocking filaments is in agreement with the simulation results and

further confirm that our strategy based on Li$^+$ flux homogenization is successful. In fact, the value is very close to the 2mA cm$^{-2}$ requirement of practical SSBs where 1C charge is carried out at an area capacity of 2 mAh cm$^{-2}$. To evaluate the sustainability of the interface, we also performed a long cyclic test. As shown in **Figure 6(g)**, at a high current density of 0.8 mA cm$^{-2}$, the H.L.-modified LLZO can stably withstand the electroplating and stripping of Li for over 1000 hours whereas the Au-modified and pristine LLZO destabilize during the initial several cycles. To further confirm the effectiveness of our strategy under practical situations, we assemble full cells using LFP and NCM523 as electrodes. In both cases, the SSBs are assembled using plasticized cathodes developed in our previous work. As shown in **Figure 6(h)-(k)**, both SSBs display good cyclic stability. In the case of LFP, the specific capacity reaches and maintains ~120 mAh g$^{-1}$ for over 300 cycles at 0.5C and a relatively high area capacity of 0.4 mAh cm$^{-2}$. For NCM523, due to the more aggressive cathode chemistry, though cyclic stability is slightly less, it is still maintained >90% of the initial value after 100 cycles. These electrochemical results further confirm that our homogenizing strategy is highly effective.

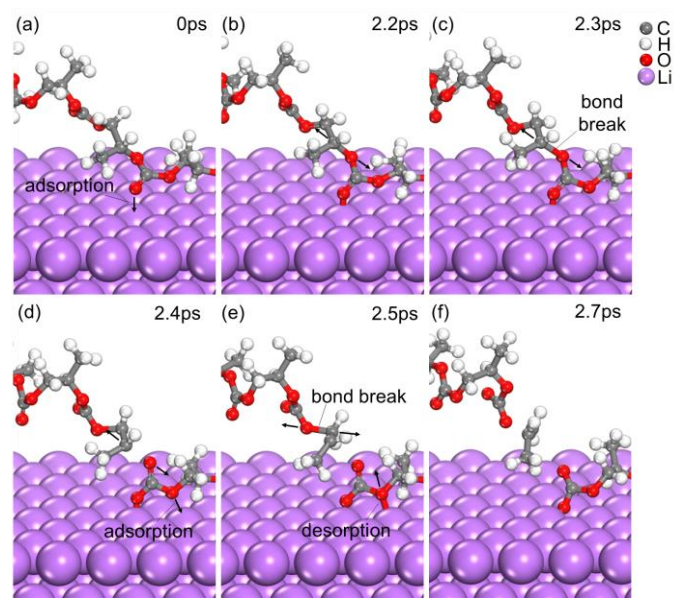

**Figure 7** Snapshots of molecular dynamics simulations of the de-polymerization reaction of PPC on Li at 500K.at (a) 0ps, (b) 2.2ps, (c) 2.3ps, (d) 2.4ps, (e) 2.5p,s and (f) 2.7ps.

To reveal the molecular origin of the excellent interfacial compatibility, the low resistance, and the conformity between the PPC-based H.L. and Li, we carried out ab-initio molecular dynamics simulations on the Li|PPC interface at 500K. We specifically chose such a high temperature to enhance the sampling of rare events. Interestingly, despite the relatively short simulation time, we were able to observe several critical processes during the de-polymerization reaction. As shown in **Figure 7(a)** and **(b)**, during the initial stage, the =O is absorbed on the Li surface followed by a strong stretching and break of the -O-C- bond, see **Figure 7(c)-(d)**. This creates two molecular fragments and decreases the molecular weight of the polymer. Due to the instability of the molecular fragments, one of them follows another decomposition as shown in **Figure 7(d)-(f)**. These fragments with low molecular weight may serve as a plasticizer and increases the ionic conductivity. This also explains the low interfacial resistance between PPC and LLZO. [52] In fact, a number of recent studies have shown that without any plasticizer, the interfacial resistance between solid-polymer electrolyte, *e.g.*, polyethylene oxide and ceramics could be as high as ~16kΩ cm$^2$. [12, 53, 54] The inclusion of a loose-binding Li$^+$ solvent, in this propylene carbonate and low-MW PPC, near the ceramic interface may facilitate Li$^+$ dissolution from the ceramic and therefore lead to a lower interfacial resistance. [55] Therefore, the superiority of PPC as an interfacial layer lies not only in its homogenous nature but also in its unique reaction mechanism with Li, which should be considered in future design of dendrite-proof interphases.

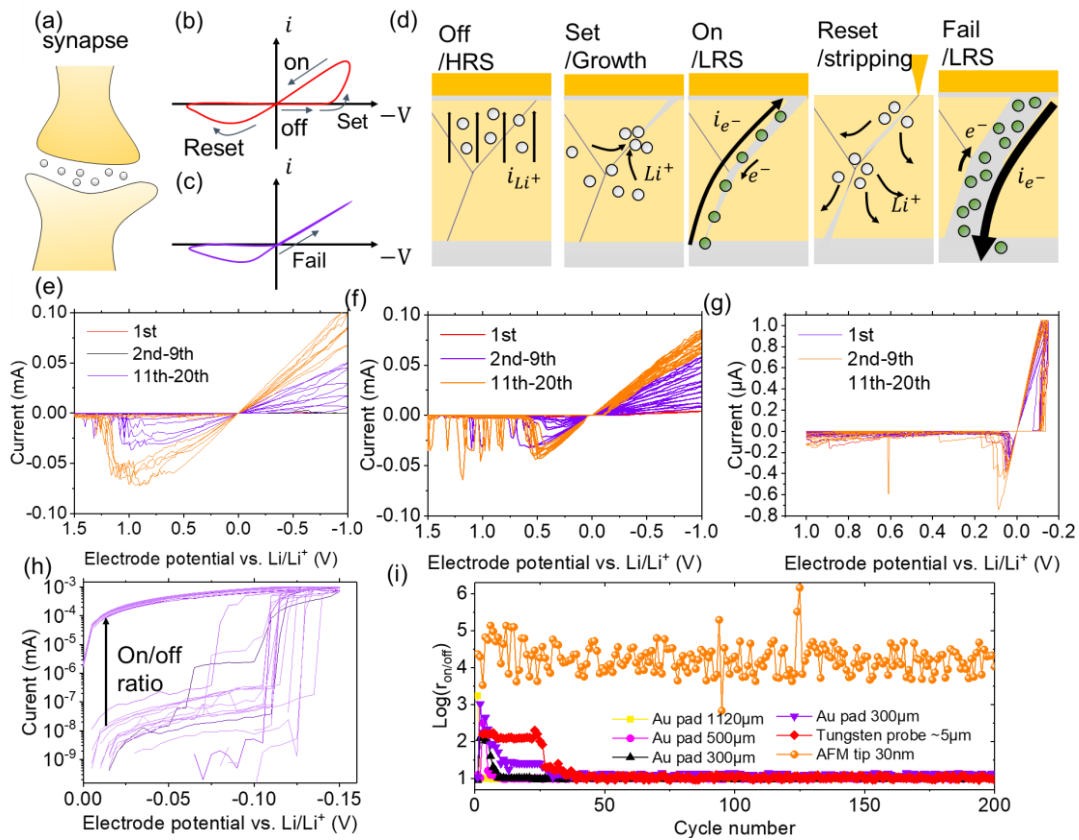

**Figure 8** Illustration of (a) a biological synapse and (b) (c) the typical response of a memristor. (d) The corresponding internal physical processes with cyclic electric bias. I-V curves of Li|LLZO|WE memristors with WE being (e) an Au pad with a 500μm-side length and (f) (g) an AFM tip. (g) The I-V curve in log scale for (g). (i) The variation of the on/off ratio with respect to cycle number for Li|LLZO|WE memristors with WE with different sizes. The set and reset potential limits are -0.2V and 1V, respectively.

Beyond the initial formation process, we also study the post-filament-penetration electrochemical kinetics of LLZO in hope of finding potential applications beyond SSBs. Interestingly, the formation of the metallic filament displays certain reversibility, *i.e.*, when a reverse bias is applied, the metallic filament tends to be absorbed by the electrolyte and the short circuit is reversed. In fact, such a phenomenon has been observed in a number of recent works. Wang *et al.* carried out *in-situ* neutron depth profiling of the Li|LLZO interface and observed "dynamic short circuiting" where during lithium plating of a half-cell, the short circuit can be temporarily eliminated. [56] Krauskopf and co-workers further confirmed such a dynamic process by carrying out galvanostatic electrochemical impedance spectroscopy. [37] A number of other recent studies also support their observations and all of these results all point to the fact that the short circuit in lithium SSBs, to some extent, can be reversed. [37, 57] We take a step further and show that the filament growth in LLZO follows a memristive behavior and the reversibility is highly dependent on the size of the electrode. Such memristive characteristics share close resemblance with biological synapse and is the key to neuromorphic computing and non-volatile memory devices, as illustrated in **Figure 8(a)**. [58] **Figure 8 (b)** and **(c)** illustrates the typical I-V curve of a Li|LLZO|WE cell. When a negative bias (*vs.* Li/Li$^+$) is applied, the Li$^+$ ions are drawn from LLZO towards the WE and are being reduced to form metallic lithium as shown in **Figure 8(d)**. During this stage, the current is controlled by the ionic conductivity of LLZO and is not measurable with the current nano-electrochemical measurement setup. This stage corresponds to the off state or the high resistance state (HRS) of an asymmetrical memristor. When the negative bias becomes larger, the lithium filament starts to grow and penetrates the LLZO pellet to form an electronically conductive path. Meanwhile, the nominal resistance becomes orders of magnitude higher. Such a transition corresponds to 'setting' a memristor to its low resistance state (LRS) or the "on" state. Depending on the size of the electrode, the reversibility of such a process can be varied. As illustrated in the fourth panel of **Figure 8(d)**, if the size of the electrode is too large and the set bias is too high, the filament becomes too thick and the electrical potential is offset by the high electron conductivity and there is not enough driving force to strip lithium away from the filament so as to break it for high resistance. Therefore, the memristor fails and cannot go back to its 'off' states. We prepared a number of WEs with different sizes as shown in the inset of **Figure 2(f)**. The Typical cyclic curves of the Li|LLZO|WE cells are shown in **Figure 8(e)-(g)**. In **Figure 8(e)**, the side length of the WE is 500μm which resembles the macroscopic case. During the initial cycle, we indeed saw and onset of the memristor. However, the ratio between the HRS and LRS state (the on/off ratio) is relatively small and did not go beyond 10. Also, after being

"set" for the first time, it can hardly be "reset". This is in agreement with most macroscopic observations where the short circuit is detrimental to an SSB and can hardly be fully reversed. [59] However, when the size of the WEs shrinks to nanoscale, the reversibility becomes much better. **Figure 8(f)** shows the results of using the c-AFM tip as WE. The "set" bias stabilizes at ~-0.25V after several cycles and reset of the memristor is always successful. In fact, the on/off ratio within an operating voltage window of ±0.25V kept almost constant after the first 10 cycles. By limiting the set and reset voltages to -0.2V and +1V, the memristive switching performance of the Li|LLZO|tip device is maximized. As shown in **Figure 8(g)**, extreme stability for over 200 cycles is achieved. **Figure 8(h)** gathers the cyclic stability of the on/off ratio *vs.* electrode sizes. Only at nanoscale, stable memristive switching can be achieved. The memristor based on the nanoelectrode displays a rather high on/off ratio of $10^5$ for 200 cycles. This may open up new opportunities for using lithium solid electrolytes in non-volatile memory and neuromorphic computing, *i.e.,* a new computing architecture beyond von Neumann, [58] and convert the notorious dendrite growth issue into something useful.

**Conclusions**

We develop an *in-situ* nano-electrochemical characterization technique based on c-AFM to reveal the nanoscopic origin of filament growth at the electrode-electrolyte interfaces in SSBs and use it to guide the design of a highly efficient filament-proof interphase. Significant nano-inhomogeneity involving fluctuations of elastic modulus and current density at the interfaces between polycrystalline LLZO and Li are identified as the major source of instability towards lithium penetration and short circuit. A highly-conductive and conformal homogenizing layer is designed to smooth out the nano-inhomogeneity and avoid the weak spots. A high CCD of 1.8 mA cm$^{-2}$ and a low interfacial resistance of 14 Ohm cm$^2$ is achieved, approaching the practical requirement of SSBs. Full SSLMB cells are demonstrated using LFP and NCM523 as cathodes and show excellent stability for up to 300 cycles. Beyond energy applications, highly stable reversible memristive behavior of lithium filament is found at nanoscale. A model memristor with a high on/off ratio of 10$^5$ is demonstrated which can be cycled for over 200 times. The interfacial strategy proposed in the current work is highly efficient in preventing lithium filament growth in SSBs and may serve as a baseline for future design of interfaces for SSBs. The nano-electrochemical characterization technique developed here not only provides insights into the nanoscopic electrochemical processes in SSBs but also opens up exciting opportunities for solid electrolyte beyond energy applications.

**Experimental and Computational Methods**

*Sample preparation.* The composition is $Li_{6.5}La_3Zr_{1.5}Ta_{0.5}O_{12}$, where the Ta doping helps stabilize the cubic phase. It is named as LLZO throughout the article for simplicity despite the Ta dopant. The solid electrolyte was synthesized via a conventional solid-state reaction where LiOH H$_2$O (≥ 99.0%, Sigma-Aldrich), La$_2$O$_3$ (99%, Sigma-Aldrich), Ta$_2$O$_5$ (≥ 99.99%, Ourchem), and ZrO$_2$ (<100 nm, Sigma-Aldrich) are used as starting materials. [60] After weighing stoichiometric amount the starting materials with 10% Li-excess and wet ball-milling with isopropanol, the mixed power was dried and sintered at 900 °C in air for 12 h followed by pelletization with another 10 wt.% LiOH H$_2$O added. Finally, the green pellets were sintered at 1140 °C for 16 h in MgO crucibles which were covered with a lid. Mother powder was added on top of the pellet to minimize the Li loss. The prepared disks were then sanded down to the thickness of 300μm. ~50nm thick Au was then deposited on one side of the pellet and a lithium foil was attached to the same side as the counter and the reference electrode followed by melting at 250 °C. This step ensures a good contact of the counter and the reference electrode.

*AFM and nanoelectrochemical measurements.* The topography and AFM current-voltage (I-V) curves were measured with a Benyuan system (CSPM5500, China) in a glove box (O$_2$ and H$_2$O <1ppm) with Keithly 2400 sourcemeter as the electrochemical measurement unit. All metal Pt probes with tip radius of 20nm (25PT300B, Rocky Mountain nanotechnology, USA) were used for CP-AFM measurement. The modulus and SKPM measurement were performed with a Demension Icon system (NANOSCOPE V7-B, Bruker, USA). The absolute value of the modulus is calibrated by matching the value of the grain interior to indentation results on LLZO grains. [61] Pt coated Si probes with a tip radius of 20nm (SCM-PIT-75, 75kHz, 2.8N m$^{-1}$) were used. Topography images and modulus images or SKPM images were taken simultaneously at a scan rate of 1Hz. During the nanoelectrochemical measurements, the current was limited to 1mA and the voltage to 10V to avoid damage to the c-AFM tip.

*Macroscopic electrochemical tests:* The H.L. was fabricated by dissolving PPC ($M_w$ = 50,000, Sigma-Aldrich) and lithium bis(trifluoromethane)-sulfonimide (LiTFSI, Sigma-Aldrich) with a weight ratio of 8:2 in N,N-dimethylformamide (Sigma-Aldrich) followed by coating the solution on LLZO pellets. The film was then vacuum dried at 80 °C for 24 h. After that, lithium foils were attached to both sides of H.L.-coated LLZO pellets and heated at 80 °C for 2h. The CCD were estimated by constant current measurement of Li|LLZO|Li cells with increasing current densities from 0 to 3mA cm$^{-2}$. The tests were carried out at 25 °C. For comparison, the cells without interlayer and with Au interlayer were fabricated similarly except that Li was melted on the pellet at 250 °C to achieve better contact. For the Au-coated case, ~50nm thick Au was deposited on LLZO. For full cell tests, LFP and NCM523 cathodes were fabricated following our previous work, by coating a N-methylpyrrolidone slurry conatining the cathode powder, the poly(vinylidene fluoride) binder, conductive carbon, succinonitrile, and LiTFSI with a weight ratio of 5:1:1:2:1 on a LLZO pellet followed by vacuum drying at 80 °C for 24 h. [60] The charge and discharge tests of the LFP and the NCM523 cells were carried out by applying constant currents to the cells. The cutoff voltages were of 2.5-4.2V and 2.8-4.2V, respectively. The testing temperatures were 40 °C for the LFP cell to achieve better rate-capability and larger cycle numbers. For NCM523 cells, the tests were carried out at 30 °C.

*Ab-initio molecular dynamics simulations: Ab-initio* molecular dynamics simulations were carried out using the CASTEP plane-wave density functional theory code. [62] The QC5 set of pseudopotential was used with a relatively low energy cutoff of 340 meV.

The calculation was spin polarized. The brillouin zone was sampled on the gamma point. The temperature was controlled at 500K using a Nosé-Hoover thermostat. [63] Such a temperature was chosen to speed up the simulation and to enhance sampling. The time step for the ionic motion was set to 1fs. The simulation box was built by including a Li surface, a linear PPC with 5 molecular units, and a vacuum layer of 15Å. The detailed model is shown in **Figure S4**.

*Finite element simulations:* The Li$^+$ flux distribution in the solid electrolyte and the H.L. was simulated by solving the electrostatics in a polycrystalline LLZO as detailed in section 4 of SI. The grain distribution was generated using a Voronoi tessellation. [64] The average size of each grain is ~10μm. The diameter of grain boundaries was set to 5nm. The detailed geometry and the mesh used in the simulation are shown in **Figure S5**. Constant potentials of 0.1V and 0V were used as boundary conditions on the two sides of the electrolyte. The materials constants used in the simulation are listed in **Table S1**. For the case of H.L.-coated LLZO, two H.L. with a thickness of 10μm each were attached to both sides of the LLZO. Other parameters were kept the same.

*Calculation of the contact area.* The contact area between LLZO and the AFM-tip was calculated using the following relation derived from Herzian contact mechanics: [65]

$$A = \pi(\frac{3P_{eff}}{4E^*}r)^{2/3}$$

where *r* is the tip radius, $P_{eff}$ is the tip pressure force, and $E^*$ is the effective modulus of LLZO, which we estimate to be ~60GPa. [66] The tip radius is about 20 nm and the tip pressure is about 400 nN calculated from our force curve measurement. From this calculation we assess the contact area in our studies to be ~30 nm$^2$.

# References

[1] J. W. Choi, D. Aurbach, *Nat. Rev. Mater.* 2016, 1, 1.
[2] D. Lin, Y. Liu, Y. Cui, *Nat. Nanotechnol.* 2017, 12, 194.
[3] B. Han, D. Xu, S. S. Chi, D. He, Z. Zhang, L. Du, M. Gu, C. Wang, H. Meng, K. Xu, *Adv. Mater.* 2020, 2004793.
[4] Q. Pang, X. Liang, C. Y. Kwok, L. F. Nazar, *Nat. Energy* 2016, 1, 1.
[5] J. Liu, Z. Bao, Y. Cui, E. J. Dufek, J. B. Goodenough, P. Khalifah, Q. Li, B. Y. Liaw, P. Liu, A. Manthiram, *Nat. Energy* 2019, 4, 180.
[6] Z. Yu, H. Wang, X. Kong, W. Huang, Y. Tsao, D. G. Mackanic, K. Wang, X. Wang, W. Huang, S. Choudhury, *Nat. Energy* 2020, 5, 526.
[7] S. Chen, J. Zheng, D. Mei, K. S. Han, M. H. Engelhard, W. Zhao, W. Xu, J. Liu, J. G. Zhang, *Adv. Mater.* 2018, 30, 1706102.
[8] D. Kang, N. Hart, J. Koh, L. Ma, W. Liang, J. Xu, S. Sardar, J. P. Lemmon, *Energy Storage Mater.* 2020, 24, 618.
[9] Q. Yang, W. Li, C. Dong, Y. Ma, Y. Yin, Q. Wu, Z. Xu, W. Ma, C. Fan, K. Sun, *J. Energy Chem.* 2020, 42, 83.
[10] J. Yi, J. Chen, Z. Yang, Y. Dai, W. Li, J. Cui, F. Ciucci, Z. Lu, C. Yang, *Adv. Energy Mater.* 2019, 9, 1901796.
[11] J. Janek, W. G. Zeier, *Nat. Energy* 2016, 1, 1.
[12] Z. Zou, Y. Li, Z. Lu, D. Wang, Y. Cui, B. Guo, Y. Li, X. Liang, J. Feng, H. Li, *Chem. Rev.* 2020, 120, 4169.
[13] H. Liu, X.-B. Cheng, J.-Q. Huang, H. Yuan, Y. Lu, C. Yan, G.-L. Zhu, R. Xu, C.-Z. Zhao, L.-P. Hou, *ACS Energy Lett.* 2020, 5, 833.
[14] R. Murugan, V. Thangadurai, W. Weppner, *Angew. Chem. Int. Ed.* 2007, 46, 7778.
[15] N. Zhao, W. Khokhar, Z. Bi, C. Shi, X. Guo, L.-Z. Fan, C.-W. Nan, *Joule* 2019, 3, 1190.
[16] C. Wang, K. Fu, S. P. Kammampata, D. W. McOwen, A. J. Samson, L. Zhang, G. T. Hitz, A. M. Nolan, E. D. Wachsman, Y. Mo, *Chem. Rev.* 2020.
[17] C. Monroe, J. Newman, *J. Electrochem. Soc.* 2005, 152, A396.
[18] Y. Ren, Y. Shen, Y. Lin, C.-W. Nan, *Electrochem. Commun.* 2015, 57, 27.
[19] X. Han, Y. Gong, K. K. Fu, X. He, G. T. Hitz, J. Dai, A. Pearse, B. Liu, H. Wang, G. Rubloff, *Nat. Mater.* 2017, 16, 572.
[20] C. Wang, Y. Gong, B. Liu, K. Fu, Y. Yao, E. Hitz, Y. Li, J. Dai, S. Xu, W. Luo, *Nano Lett.* 2017, 17, 565.
[21] A. Sharafi, E. Kazyak, A. L. Davis, S. Yu, T. Thompson, D. J. Siegel, N. P. Dasgupta, J. Sakamoto, *Chem. Mater.* 2017, 29, 7961.
[22] A. Sharafi, H. M. Meyer, J. Nanda, J. Wolfenstine, J. Sakamoto, *J. Power Sources* 2016, 302, 135.
[23] F. Han, A. S. Westover, J. Yue, X. Fan, F. Wang, M. Chi, D. N. Leonard, N. J. Dudney, H. Wang, C. Wang, *Nat. Energy* 2019, 4, 187.
[24] S. Yu, D. J. Siegel, *ACS Appl. Mater. Interfaces* 2018, 10, 38151.

[25] P. Barai, K. Higa, A. T. Ngo, L. A. Curtiss, V. Srinivasan, *J. Electrochem. Soc.* 2019, 166, A1752.
[26] J. F. Nonemacher, Y. Arinicheva, G. Yan, M. Finsterbusch, M. Krüger, J. Malzbender, *Journal of the European Ceramic Society* 2020.
[27] L. Porz, T. Swamy, B. W. Sheldon, D. Rettenwander, T. Frömling, H. L. Thaman, S. Berendts, R. Uecker, W. C. Carter, Y. M. Chiang, *Adv. Energy Mater.* 2017, 7, 1701003.
[28] J. Wen, Y. Huang, J. Duan, Y. Wu, W. Luo, L. Zhou, C. Hu, L. Huang, X. Zheng, W. Yang, *ACS Nano* 2019, 13, 14549.
[29] X. Fu, T. Wang, W. Shen, M. Jiang, Y. Wang, Q. Dai, D. Wang, Z. Qiu, Y. Zhang, K. Deng, *Adv. Mater.* 2020, 2000575.
[30] B. Xu, W. Li, H. Duan, H. Wang, Y. Guo, H. Li, H. Liu, *J. Power Sources* 2017, 354, 68.
[31] H. Huo, Y. Chen, R. Li, N. Zhao, J. Luo, J. G. P. da Silva, R. Mücke, P. Kaghazchi, X. Guo, X. Sun, *Energy Environ. Sci.* 2020, 13, 127.
[32] H. Huo, J. Liang, N. Zhao, X. Li, X. Lin, Y. Zhao, K. Adair, R. Li, X. Guo, X. Sun, *ACS Energy Lett.* 2020.
[33] W. Feng, X. Dong, X. Zhang, Z. Lai, P. Li, C. Wang, Y. Wang, Y. Xia, *Angew. Chem. Int. Ed.* 2020, 59, 5346.
[34] J. Meng, Y. Zhang, X. Zhou, M. Lei, C. Li, *Nat. Commun.* 2020, 11, 1.
[35] J. J. Yang, D. B. Strukov, D. R. Stewart, *Nat. Nanotechnol.* 2013, 8, 13.
[36] Q. Xia, J. J. Yang, *Nat. Mater.* 2019, 18, 309.
[37] T. Krauskopf, R. Dippel, H. Hartmann, K. Peppler, B. Mogwitz, F. H. Richter, W. G. Zeier, J. Janek, *Joule* 2019, 3, 2030.
[38] Y. Song, L. Yang, W. Zhao, Z. Wang, Y. Zhao, Z. Wang, Q. Zhao, H. Liu, F. Pan, *Adv. Energy Mater.* 2019, 9, 1900671.
[39] Y. Song, L. Yang, L. Tao, Q. Zhao, Z. Wang, Y. Cui, H. Liu, Y. Lin, F. Pan, *J. Mater. Chem. A* 2019, 7, 22898.
[40] H.-K. Tian, B. Xu, Y. Qi, *J. Power Sources* 2018, 392, 79.
[41] J.-F. Wu, X. Guo, *Phys. Chem. Chem. Phys.* 2017, 19, 5880.
[42] N. Goswami, R. Kant, *Journal of Electroanalytical Chemistry* 2019, 835, 227.
[43] M. Nonnenmacher, M. o'Boyle, H. K. Wickramasinghe, *Applied physics letters* 1991, 58, 2921.
[44] H. Shiiba, N. Zettsu, M. Yamashita, H. Onodera, R. Jalem, M. Nakayama, K. Teshima, *The Journal of Physical Chemistry C* 2018, 122, 21755.
[45] S. Yu, D. J. Siegel, *Chem. Mater.* 2017, 29, 9639.
[46] W. Li, S. Cohen, K. Gartsman, R. Caballero, P. Van Huth, R. Popovitz-Biro, D. Cahen, *Solar energy materials and solar cells* 2012, 98, 78.
[47] J. Zhang, J. Zhao, L. Yue, Q. Wang, J. Chai, Z. Liu, X. Zhou, H. Li, Y. Guo, G. Cui, *Adv. Energy Mater.* 2015, 5, 1501082.
[48] J. Zhang, X. Zang, H. Wen, T. Dong, J. Chai, Y. Li, B. Chen, J. Zhao, S. Dong, J. Ma, *J. Mater. Chem. A* 2017, 5, 4940.
[49] Y. Li, F. Ding, Z. Xu, L. Sang, L. Ren, W. Ni, X. Liu, *J. Power Sources* 2018, 397, 95.

[50] B. Commarieu, A. Paolella, S. Collin-Martin, C. Gagnon, A. Vijh, A. Guerfi, K. Zaghib, *J. Power Sources* 2019, 436, 226852.
[51] M.-x. Jing, H. Yang, H. Chen, S. Hua, B.-w. Ju, Q. Zhou, F.-y. Tu, X.-q. Shen, S.-b. Qin, *SN Applied Sciences* 2019, 1, 205.
[52] C. Wang, H. Zhang, J. Li, J. Chai, S. Dong, G. Cui, *J. Power Sources* 2018, 397, 157.
[53] D. Brogioli, F. Langer, R. Kun, F. La Mantia, *ACS Appl. Mater. Interfaces* 2019, 11, 11999.
[54] A. Gupta, J. Sakamoto, *Electrochemical Society Interface* 2019, 28, 63.
[55] X. C. Chen, X. Liu, A. Samuthira Pandian, K. Lou, F. M. Delnick, N. J. Dudney, *ACS Energy Lett.* 2019, 4, 1080.
[56] C. Wang, Y. Gong, J. Dai, L. Zhang, H. Xie, G. Pastel, B. Liu, E. Wachsman, H. Wang, L. Hu, *J. Am. Chem. Soc.* 2017, 139, 14257.
[57] W. Ping, C. Wang, Z. Lin, E. Hitz, C. Yang, H. Wang, L. Hu, *Adv. Energy Mater.* 2020, 2000702.
[58] Z. Wang, S. Joshi, S. E. Savel'ev, H. Jiang, R. Midya, P. Lin, M. Hu, N. Ge, J. P. Strachan, Z. Li, *Nat. Mater.* 2017, 16, 101.
[59] D. Cao, X. Sun, Q. Li, A. Natan, P. Xiang, H. Zhu, *Matter* 2020.
[60] Z. Lu, J. Yu, J. Wu, M. B. Effat, S. C. Kwok, Y. Lyu, M. M. Yuen, F. Ciucci, *Energy Storage Mater.* 2019, 18, 311.
[61] J. E. Ni, E. D. Case, J. S. Sakamoto, E. Rangasamy, J. B. Wolfenstine, *J. Mater. Sc.* 2012, 47, 7978.
[62] M. Segall, P. J. Lindan, M. a. Probert, C. J. Pickard, P. J. Hasnip, S. Clark, M. Payne, *Journal of physics: condensed matter* 2002, 14, 2717.
[63] D. J. Evans, B. L. Holian, *The Journal of chemical physics* 1985, 83, 4069.
[64] Q. Du, V. Faber, M. Gunzburger, *SIAM review* 1999, 41, 637.
[65] V. B. Engelkes, J. M. Beebe, C. D. Frisbie, *The Journal of Physical Chemistry B* 2005, 109, 16801.
[66] S. Yu, R. D. Schmidt, R. Garcia-Mendez, E. Herbert, N. J. Dudney, J. B. Wolfenstine, J. Sakamoto, D. J. Siegel, *Chem. Mater.* 2016, 28, 197.

**Acknowledgements**

This work was supported by the Basic Research Program of Shenzhen (No. JCYJ20190812161409163), the Basic and Applied Basic Research Program of Guangdong Province (No. 2019A1515110531), the SIAT Innovation Program for Excellent Young Researchers.

TOC

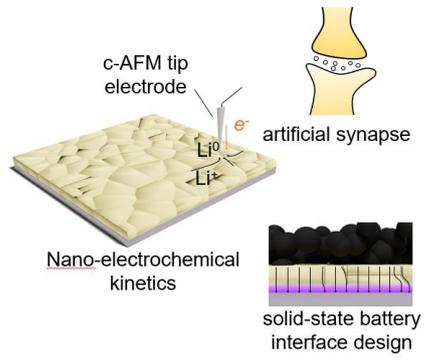